\documentclass[12pt]{article}
\usepackage{latexsym}
\usepackage{amsmath}
\usepackage{amssymb}
\usepackage{amsthm}
\usepackage{dsfont}
\usepackage{longtable}
\usepackage{color}
\usepackage{verbatim}
\usepackage[sort&compress,numbers]{natbib}
\usepackage{graphicx}
\usepackage{epstopdf}
\usepackage{float}
\usepackage{wrapfig}

\usepackage[breaklinks]{hyperref}
\hypersetup{
    colorlinks,
    citecolor=black,
    filecolor=black,
    linkcolor=black,
    urlcolor=blue
}

\newcommand {\dd}[2] {\frac {\partial {#1} }{\partial {#2}}}

\newcommand{\Tr}{\mathop{\mathrm{Tr}}\nolimits}

\theoremstyle{definition}

\makeatletter
\@addtoreset{remark}{utv}
\@addtoreset{remark}{theorem}
\makeatother

\allowdisplaybreaks
\title{Exact expressions for the intercepts of $r$-particle momentum correlation functions in $\mu$-Bose gas model}
\author{A.M. Gavrilik and Yu.A. Mishchenko}
\date{}

\begin{document}
\maketitle

\begin{abstract}
Recently, the deformed $\mu$-Bose gas model based on so called $\mu$-deformed oscillators 
was proposed. For that model, the intercepts of $r$-particle momentum correlation functions 
(correlation functions at coinciding momenta of particles) were treated for $r=2,3$, within 
certain order of approximation in $\mu$. In this work, we derive for $\mu$-Bose gas model 
the {\it exact} expressions for $r$-particle correlation function intercepts, for all $r$, 
through  Lerch transcendent and find their asymptotics (as functions of $\mu$). For 
2-, 3-particle intercepts and deformed distribution function $\langle a^\dag_k a_k\rangle$ 
the dependence on particles momentum is presented graphically.
\end{abstract}
{\bf Keywords}: deformed Bose gas model, deformed oscillators, non-standard statistics, 
intercepts of correlation functions, Lerch transcendent.
 \newline
{\bf PACS}: 05.30.Jp; 05.90.+m; 11.10.Lm; 25.75.Gz; 02.30.Gp.

\section{Introduction}

The deformed $\mu$-Bose gas model~\cite{GR_EPJA} is a particular implementation of such an efficient in nonlinear quantum physics tools as deformed oscillators (for their definition, diverse models and some applications, see e.g.~\cite{Bonatsos}). The latter provide a generalization of the quantum harmonic oscillator and are often used in effective descriptions of various nontrivial properties of nonlinear systems and/or phenomena. Due to  an effective representation in terms of deformed oscillator(s) we may come to a more simple system whose behavior approximates the behavior of the initial one. Main properties of a deformed oscillator and its distinction from the standard quantum harmonic oscillator are characterized by the corresponding deformation structure function~$\phi(N)$ (see~\cite{Meljanac,Bonatsos,Daskaloyannis_1991} for its notion). Depending on~$\phi(N)$ one distinguishes many types of deformed oscillators. Say, the one used in the elaboration of $\mu$-Bose gas model is called $\mu$-oscillator and was introduced in~\cite{Janussis}. Besides this rather exotic $\mu$-deformed oscillator (for which, $\phi(N)=\frac{N}{1+\mu N}$) there are such well-known $q$-deformed oscillators as Arik-Coon~\cite{Arik_Coon} and Biedenharn-Macfarlane~\cite{Biedenharn,Macfarlane} ones, the $p,q$-deformed oscillator defined in~\cite{pq-Oscillator} and also the $q$-deformed Tamm-Dancoff oscillator~\cite{Odaka-Kamefuchi(TD_Osc),Chaturvedi-Jagannathan(TD_Osc)}. Let us note that the latter four deformed oscillator models belong to the class of Fibonacci oscillators~\cite{Arik_1992} (i.e. such that their structure function $\phi_n\equiv \phi(n)$ satisfies Fibonacci-like three-term recurrence relation with constant coefficients). Unlike, the $\mu$-oscillator represents the class of so-called quasi-Fibonacci deformed oscillators treated in~\cite{Gavrilik_QuasiFibonacci} and that implies some peculiarities in concrete applications.

Besides the $\mu$-Bose gas model there exist many other deformed Bose gas models in which deformed oscillators were utilized. To list a few, the $q$-Bose gas model was studied in~\cite{Anchishkin2000} (see also the references therein), $p,q$-Bose gas model was investigated in~\cite{Adamska,Gavrilik_Sigma} and in some other works. Thermodynamical aspects of deformed Bose gases were studied e.g. in~\cite{Lavagno_Therm,Lavagno_Thermostat,Algin_2002}. It is worth to mention that deformed oscillators were also effectively used for modelling a gas of (nonpointlike, composite) particles or quasiparticles~\cite{Avan,Bagheri_Harouni,Liu}, in the study of Bose-Einstein condensation~\cite{Avancini_Marinelli} and of phonon spectrum in ${}^4He$~\cite{R-Monteiro}, for the description of interacting particles systems~\cite{Scarfone_Interact}. As it was shown in~\cite{GKM,GKM2}, it is possible to realize  composite bosons in terms of deformed oscillators (deformed bosons), and to relate the inter-component entanglement just with the parameter of deformation~\cite{GM_Entang}. In the context of $n$-particle correlations, deformed oscillators were applied in~\cite{Anchishkin2000,Adamska,Gavrilik_Sigma}, and the corresponding intercepts of correlation functions (along with their asymptotics) were obtained as functions of deformation parameter.

In the preceding paper~\cite{GR_EPJA} the approximate expressions to certain order in $\mu$ were obtained for the intercepts of the 2nd
and 3rd order momentum correlation functions in the deformed $\mu$-Bose gas model, and for the corresponding asymptotics the exact expressions were found. It turns out that the results for the $r$th order intercepts can be obtained in exact form and this constitutes main subject of the present work. In fact, we derive the corresponding exact formulae given in terms of Lerch Phi-function (Lerch transcendent) for the general case of $r$-particle correlation intercept and also illustrate some of them graphically, with special attention to the particular cases $r=2$ and $r=3$. One can hope that present results and further development $\mu$-Bose gas model, say, concerning thermodynamical aspects will find effective applications to diverse physical systems.

\section{Derivation of the exact expressions for intercepts of the correlation functions in $\mu$-Bose gas model}\label{sec2}

Our treatment concerns deformed $\mu$-Bose gas model as the secondary quantized system of specially deformed bosons. That means that each mode~$\bf k$ in the deformed Bose gas is described by the deformed oscillator with the creation/annihilation operators $a^\dag_{\bf k}$, $a_{\bf k}$, and the number operator $N_{\bf k}$. The defining relations for such a system of deformed oscillators (deformed bosons) are given as
\begin{align}
&a^\dag_{\bf k}a_{\bf k} = \phi(N_{\bf k}),\label{def1}\\
&[a_{\bf k},a^\dag_{\bf k'}] = \delta_{\bf kk'} \bigl(\phi(N_{\bf k}+1)-\phi(N_{\bf k})\bigr),\quad [a_{\bf k},a_{\bf k'}]=0,\label{def2}\\
&[N_{\bf k},a^\dag_{\bf k'}] = \delta_{\bf kk'} a^\dag_{\bf k},\quad [N_{\bf k},a_{\bf k'}] = -\delta_{\bf kk'} a_{\bf k}\label{def3},
\end{align}
where the concept of structure function~$\phi(N)$ is used. As seen, we deal with the system of deformed oscillators which are mode independent i.e. the operators corresponding to deformed oscillators in different modes commute.

Like in~\cite{GR_EPJA}, and similarly to the corresponding nondeformed analogs~\cite{Wiedemann1999145}, the concerned intercepts of the $r$th order momentum correlation function at a given momentum~$\bf k$ are defined as
\begin{equation}\label{lambda_def}
\lambda^{(r)}({\bf k}) = \frac{\langle (a^\dag_{\bf k})^r (a_{\bf k})^r \rangle}{\langle a^\dag_{\bf k} a_{\bf k}\rangle^r}-1.
\end{equation}
Symbol $\langle...\rangle$ denotes usual statistical average for a system with Hamiltonian~$H$:
\begin{equation}\label{<F>}
\langle F\rangle = \frac{\Tr F\exp(-\beta H)}{\Tr \exp(-\beta H)},
\end{equation}
where $\beta=(k_B T)^{-1}$, $k_B$ is Boltzmann constant.
Using (\ref{def1}) and (\ref{def3}) the formula~(\ref{lambda_def}) for the intercepts~$\lambda^{(r)}({\bf k})$ can be rewritten in terms of the structure function~$\phi(N)$ as
\begin{equation}\label{lambda_def2}
\lambda^{(r)}({\bf k}) = \frac{\langle[N_{\bf k}][N_{\bf k}-1]\cdot...\cdot[N_{\bf k}-r+1]\rangle}{\langle [N_{\bf k}]\rangle^r}-1,
\end{equation}
where the notation $[N] = [N]_\phi \equiv \phi(N)$ for the structure function is used. Since we deal with the $\mu$-deformed Bose gas model~\cite{GR_EPJA}, for $\phi(N)$ we take the structure function of $\mu$-deformed oscillator:
\begin{equation*}
\phi_\mu(N) \equiv [N]_\mu = \frac{N}{1+\mu N},\quad \mu>0.
\end{equation*}
The intercepts~(\ref{lambda_def2}) depend on the choice of Hamiltonian $H$. We choose the Hamiltonian in the form
\begin{equation}\label{H}
H= \sum_{\bf k} \hbar\omega_{\bf k} N_{\bf k}
\end{equation}
and assume that deformed boson energy depends only on the absolute value $k=|{\bf k}|$ of the momentum. Then, using the Fock-like basis in which $N_{\bf k}|n_{\bf k}\rangle = n_{\bf k}|n_{\bf k}\rangle$, $\phi(N_{\bf k})|n_{\bf k}\rangle = \phi(n_{\bf k})|n_{\bf k}\rangle$, the mode independence, and taking into account (\ref{<F>})-(\ref{H}), for the $r$-particle correlation function intercept we obtain
\begin{equation}\label{lambda_def3}
\lambda^{(r)}_\mu(k) = \frac{(1-e^{-\beta\hbar\omega})^{-r+1}\sum_{n=0}^\infty \frac{n}{1+\mu n}\cdot...\cdot\frac{n-r+1}{1+\mu (n-r+1)} \exp\bigl[-\beta \hbar\omega n\bigr]}{\Bigl(\sum_{n=0}^\infty \frac{n}{1+\mu n} \exp\bigl[-\beta \hbar\omega n)\bigr]\Bigr)^r} - 1,\ \ \omega\!=\!\omega_{\bf k}\!=\!\omega(k).
\end{equation}
As seen, the intercept depends on $\mu$, on the momentum absolute value $k$ and temperature $T$. To proceed, we expand the product in the $n$th summand of the numerator series (in front of exponent) into the sum of simple fractions:
\begin{equation}\label{frac_expansion}
\prod_{l=0}^{r-1} \frac{n-l}{1+\mu(n-l)} = \frac{1}{\mu^r} \prod_{l=0}^{r-1} \Bigl(1-\frac{1}{1+\mu(n\!-\!l)}\Bigr) = \frac{1}{\mu^r} + \frac{1}{\mu^r} \sum_{l=0}^{r-1} \frac{A^{(r)}_l(\mu)}{1+\mu(n-l)}.
\end{equation}
The coefficients~$A^{(r)}_l(\mu)$ obey the recurrence relations:
\begin{equation}\label{expansion_coefs}
\left\{
\begin{aligned}
&A^{(r+1)}_l(\mu) = A^{(r)}_l(\mu) \Bigl(1+\frac{1}{\mu(r-l)}\Bigr),\quad l=0,...,r-1;\\
&A^{(r+1)}_r(\mu) = -1 - \sum_{l=0}^{r-1} \frac{A^{(r)}_l(\mu)}{\mu(r-l)}.
\end{aligned}
\right.
\end{equation}
These relations allow to find the coefficients $A^{(r)}_l$, to list a few:
\begin{align*}
& A^{(1)}_0(\mu) = -1;\\
& A^{(2)}_0(\mu) = -1-\frac{1}{\mu},\quad A^{(2)}_1(\mu) = -1+\frac{1}{\mu};\\
& A^{(3)}_0(\mu) = -1-\frac{3}{2\mu}-\frac{1}{2\mu^2},\quad A^{(3)}_1(\mu) = -1+\frac{1}{\mu^2},\quad A^{(3)}_2(\mu) = -1 + \frac{3}{2\mu}-\frac{1}{2\mu^2};\\
&\ldots\, .
\end{align*}
Then, in view of~(\ref{frac_expansion})-(\ref{expansion_coefs}) the infinite sum in the numerator of~(\ref{lambda_def3}) can be rewritten as
\begin{multline}\label{num_expr}
\sum_{n=0}^\infty \frac{n}{1+\mu n}\cdot...\cdot\frac{n-r+1}{1+\mu (n-r+1)} \exp\bigl[-\beta \hbar\omega n\bigr] = \frac{1}{\mu^r} \sum_{n=0}^\infty \exp\bigl[-\beta \hbar\omega n\bigr] +\\
+ \frac{1}{\mu^r} \sum_{l=0}^{r-1} A^{(r)}_l(\mu) \sum_{n=0}^\infty \frac{1}{1+\mu(n-l)}\exp\bigl[-\beta \hbar\omega n\bigr].
\end{multline}
Likewise for the sum in the denominator of~(\ref{lambda_def3}) we obtain
\begin{equation}\label{denum_expr}
\sum_{n=0}^\infty \frac{n}{1+\mu n} \exp\bigl[-\beta \hbar\omega n\bigr] = \frac{1}{\mu} \sum_{n=0}^\infty \exp\bigl[-\beta \hbar\omega n\bigr] + \frac{1}{\mu} A^{(1)}_0(\mu) \sum_{n=0}^\infty \frac{1}{1+\mu n}\exp\bigl[-\beta \hbar\omega n\bigr].
\end{equation}
As seen, the second series in the r.h.s. of~(\ref{num_expr}) and~(\ref{denum_expr}) are of similar form. Therefore in order to calculate this expression it is enough to perform summation of the generic series, that yields:
\begin{align*}
&\sum_{n=0}^\infty \frac{1}{a+b n} e^{-\beta\hbar\omega n} = \sum_{n=0}^\infty \frac{z^n}{a+b n} \equiv \frac1b \Phi(z,1,a/b) = \frac1b \Phi(e^{-\beta\hbar\omega },1,a/b),\\
&a=1-\mu l,\ \ b=\mu,\ \ l=1,...,r-1,\quad\ \  z=e^{-\beta\hbar\omega},
\end{align*}
where $\Phi$ is the Lerch transcendent, see e.g.~\cite{Gradshteyn_Ryzhik(2007)}.

As result we arrive at the following expression for the thermal average $\langle a^\dag_k a_k\rangle$
\begin{equation}\label{a^dag_a}
\langle a^\dag_k a_k\rangle = \langle\phi(N_k)\rangle =
\left\{
\begin{aligned}
&(e^{\beta \hbar\omega}-1)^{-1},\quad \mu=0,\\
&\mu^{-1} - \mu^{-2} (1-e^{-\beta \hbar\omega}) \Phi(e^{-\beta\hbar\omega},1,\mu^{-1}),\quad \mu>0,
\end{aligned}
\right.\\
\end{equation}
and the $r$th order averages (deformed analogs of $r$-particle momentum distribution)
\begin{equation}\label{a_dag^ra^r}
\langle(a^\dag_k)^r (a_k)^r\rangle = 1 + \mu^{-1} (1-e^{-\beta \hbar\omega}) \sum_{l=0}^{r-1} A^{(r)}_l(\mu) \Phi(e^{-\beta\hbar\omega},1,\mu^{-1}-l),\quad r\ge 2,
\end{equation}
which constitute one of our main results. The limit $\mu\rightarrow0$ applied to~(\ref{a_dag^ra^r}) has a peculiarity, so we calculate $\langle(a^\dag_k)^r (a_k)^r\rangle$ for $\mu=0$ separately. Starting from its definition we arrive at
\begin{equation*}
\langle(a^\dag_k)^r (a_k)^r\rangle|_{\mu=0} = (1-e^{-\beta\hbar\omega})\sum_{n=0}^\infty n(n-1)...(n-r+1) e^{-\beta\hbar\omega n}.
\end{equation*}
Using Abel's identity for the latter sum we obtain:
\begin{multline*}
(1-e^{-\beta\hbar\omega})\sum_{n=0}^\infty n(n-1)...(n-r+1) e^{-\beta\hbar\omega n} = (1-e^{-\beta\hbar\omega}) \lim_{N\rightarrow\infty} \biggl\{N(N-1)...(N-r+1) \frac{e^{-\beta\hbar\omega(N+1)}-1}{e^{-\beta\hbar\omega}-1} -\\
-\sum_{n=0}^{N-1} r\, n(n-1)...(n-r+2) \frac{e^{-\beta\hbar\omega(n+1)}-1}{e^{-\beta\hbar\omega}-1}\biggr\} = (1-e^{-\beta\hbar\omega}) \frac{r}{e^{\beta\hbar\omega}-1} \sum_{n=0}^\infty n(n-1)...(n-r+2) e^{-\beta\hbar\omega n} =\\
= \frac{r}{e^{\beta\hbar\omega}-1} \langle(a^\dag_k)^{r-1} (a_k)^{r-1}\rangle|_{\mu=0} = \ldots = \frac{r!}{(e^{\beta\hbar\omega}-1)^r}
\end{multline*}
Thus,
\begin{equation}\label{mu=0_case}
\langle(a^\dag_k)^r (a_k)^r\rangle|_{\mu=0} = \frac{r!}{(e^{\beta\hbar\omega}-1)^r}.
\end{equation}
From (\ref{a^dag_a})-(\ref{a_dag^ra^r})  and eq.(\ref{lambda_def}) we obtain our next result -- the intercept $\lambda^{(r)}_\mu(k)$ of $r$th order correlation function,
\begin{align}\label{lambda_r}
&\lambda^{(r)}_\mu(k) = \Bigl(1 + \mu^{-1} (1-e^{-\beta \hbar\omega}) \sum_{l=0}^{r-1} A^{(r)}_l(\mu) \Phi(e^{-\beta\hbar\omega},1,\mu^{-1}-l)\Bigr)\cdot \nonumber\\
&\cdot\Bigl(1 + \mu^{-1} (1-e^{-\beta \hbar\omega}) A^{(1)}_0(\mu) \Phi(e^{-\beta\hbar\omega},1,\mu^{-1})\Bigr)^{-r} - 1,\quad r=2,3,...\,,\quad \text{if}\ \mu>0,
\end{align}
Using (\ref{mu=0_case}), in the no-deformation limit
$\mu\rightarrow0$ we recover:
\begin{equation}\label{lambda_r_0}
\lambda^{(r)}_\mu(k)|_{\mu=0} = r!-1.
\end{equation}
The obtained formula (\ref{lambda_r}) presents the {\it exact general} expression for the intercepts under consideration as it covers all the orders $r\ge 2$.

\section{Series expansions in $\mu$}

In the case of small deformation parameter $\mu$ it may be of interest to have an expansion of $\lambda^{(r)}_\mu(k)$ in $\mu$. For that goal we expand the generic series present in (\ref{num_expr}) and (\ref{denum_expr}) as follows:
\begin{equation}\label{1frac}
\sum_{n=0}^\infty \frac{1}{1+\mu (n-l)} e^{-\alpha n} = \sum_{s=0}^\infty c_s(l) \mu^s,\quad \alpha = \beta\hbar\omega,\ l=0,...,r-1.
\end{equation}
The coefficients $c_s(l)$ of the expansion are found as
\begin{multline*}
c_s(l) = \frac{1}{s!} \dd{^s}{\mu^s} \sum_{n=0}^\infty \frac{1}{1+\mu (n-l)} e^{-\alpha n} \Bigr|_{\mu=0} = e^{-\alpha l} \sum_{n=0}^\infty (-1)^s (n-l)^s e^{-\alpha (n-l)} = (-1)^s e^{-\alpha l} \sum_{n=-l}^\infty n^s e^{-\alpha n} =\\
= (-1)^s e^{-\alpha l} \Bigl(-\dd{}{\alpha}\Bigr)^s \sum_{n=-l}^\infty e^{-\alpha n} = e^{-\alpha l} \Bigl(\dd{}{\alpha}\Bigr)^s \frac{e^{\alpha l}}{1-e^{-\alpha}} = \frac{1}{x^l} \Bigl(x\frac{d}{dx}\Bigr)^s \frac{x^{l+1}}{x-1},\quad x=e^\alpha.
\end{multline*}
For the derivative in the last expression we perform the following transformation ($m=l+1$):
\begin{align*}
\Bigl(x\frac{d}{dx}\Bigr)^s \frac{x^m}{x-1} &= \Bigl(x\frac{d}{dx}\Bigr)^s \frac{x^m-1}{x-1} + \Bigl(x\frac{d}{dx}\Bigr)^s \frac{1}{x-1} = \Bigl(x\frac{d}{dx}\Bigr)^s \sum_{j=0}^{m-1} x^j + \Bigl(x\frac{d}{dx}\Bigr)^s \frac{1}{x-1} =\\
&= \delta_{s0} + \sum_{j=1}^{m-1} j^s x^j + \Bigl(x\frac{d}{dx}\Bigr)^s \frac{1}{x-1}.
\end{align*}
By induction it can be checked that the latter derivative may be presented as the finite sum:
\begin{equation*}
\Bigl(x\frac{d}{dx}\Bigr)^s \frac{1}{x-1} = \sum_{j=0}^s \frac{(-1)^s}{j+1} \frac{g_s^j}{(x-1)^{j+1}}
\end{equation*}
where the coefficients $g_s^j$ satisfy the recurrence relation
\begin{equation*}
g^j_{s+1} = (j\!+\!1) (g_s^j + g_s^{j-1}),\quad g_0^0=1.
\end{equation*}
The solution of this recurrence relation is expressed through the Stirling numbers of the second kind $\bigl\{{s\atop j}\bigr\}$:
\begin{equation*}
g_s^j = (j+1)! \Bigl\{{s\!+\!1\atop j\!+\!1}\Bigr\}.
\end{equation*}
In view of the latter, the coefficients $c_s(l)$ take the form
\begin{equation*}
c_s(l) = \delta_{s0} e^{-\beta \hbar\omega l} + e^{-\beta \hbar\omega l} \sum_{j=1}^{l} j^s e^{\beta \hbar\omega j}+ e^{-\beta \hbar\omega l} (-1)^s \sum_{j=0}^s j! \Bigl\{{s\!+\!1\atop j\!+\!1}\Bigr\} \frac{1}{(e^{\beta \hbar\omega}-1)^{j+1}}.
\end{equation*}
Let us write out several first coefficients:
\begin{align*}
&c_0(0) = 1 + \frac{1}{(e^{\beta \hbar\omega}\!-\!1)},\\
&c_1(0) = - \frac{1}{(e^{\beta \hbar\omega}\!-\!1)} - \frac{1}{(e^{\beta \hbar\omega}\!-\!1)^{2}},\\
&c_2(0) = \frac{1}{(e^{\beta \hbar\omega}\!-\!1)} + \frac{3}{(e^{\beta \hbar\omega}\!-\!1)^{2}} +  \frac{2}{(e^{\beta \hbar\omega}\!-\!1)^{3}},\\
&c_3(0) = - \frac{1}{(e^{\beta \hbar\omega}\!-\!1)} - \frac{7}{(e^{\beta \hbar\omega}\!-\!1)^{2}} - \frac{12}{(e^{\beta \hbar\omega}\!-\!1)^{3}} - \frac{6}{(e^{\beta \hbar\omega}\!-\!1)^{4}},\\
&c_4(0) = \frac{1}{(e^{\beta \hbar\omega}\!-\!1)} + \frac{15}{(e^{\beta \hbar\omega}\!-\!1)^{2}} + \frac{50}{(e^{\beta \hbar\omega}\!-\!1)^{3}} + \frac{60}{(e^{\beta \hbar\omega}\!-\!1)^{4}} + \frac{24}{(e^{\beta \hbar\omega}\!-\!1)^{5}},\\
&c_5(0) = - \frac{1}{(e^{\beta \hbar\omega}\!-\!1)} - \frac{31}{(e^{\beta \hbar\omega}\!-\!1)^{2}} - \frac{180}{(e^{\beta \hbar\omega}\!-\!1)^{3}} - \frac{390}{(e^{\beta \hbar\omega}\!-\!1)^{4}} - \frac{360}{(e^{\beta \hbar\omega}\!-\!1)^{5}} - \frac{120}{(e^{\beta \hbar\omega}\!-\!1)^{6}},\\
&c_6(0) = \frac{1}{(e^{\beta \hbar\omega}\!-\!1)} + \frac{63}{(e^{\beta \hbar\omega}\!-\!1)^{2}} + \frac{602}{(e^{\beta \hbar\omega}\!-\!1)^{3}} + \frac{2100}{(e^{\beta \hbar\omega}\!-\!1)^{4}} + \frac{3360}{(e^{\beta \hbar\omega}\!-\!1)^{5}} + \frac{2520}{(e^{\beta \hbar\omega}\!-\!1)^{6}} + \frac{840}{(e^{\beta \hbar\omega}\!-\!1)^{7}}.
\end{align*}
Using the ``initial'' coefficients $c_s(0)$, all other coefficients can be determined by the formula
\begin{equation*}
c_s(l) = e^{-\beta \hbar\omega l} \sum_{j=1}^{l} j^s e^{\beta \hbar\omega j} + e^{-\beta \hbar\omega l} c_s(0).
\end{equation*}
Then the sums~(\ref{num_expr}) and~(\ref{denum_expr}), the latter being a particular case of the former, have the following Taylor expansions:
\begin{align}
&\sum_{n=0}^\infty \frac{n}{1\!+\!\mu n}\cdot...\cdot\frac{n\!-\!r\!+\!1}{1\!+\!\mu (n\!-\!r\!+\!1)} e^{-\beta \hbar\omega n} = \frac{1}{\mu^r} (1-e^{-\beta \hbar\omega})^{-1} + \frac{1}{\mu^r} \sum_{s=0}^\infty \sum_{l=0}^{r-1} A^{(r)}_l(\mu) c_s(l) \,\mu^s,\label{num_expansion}\\
&\sum_{n=0}^\infty \frac{n}{1\!+\!\mu n} e^{-\beta \hbar\omega n} = \frac{1}{\mu} (1-e^{-\beta \hbar\omega})^{-1} + \frac{1}{\mu} \sum_{s=0}^\infty A^{(1)}_0(\mu) c_s(0) \,\mu^s.\label{denum_expansion}
\end{align}
The series (\ref{num_expansion}) and (\ref{denum_expansion}) can be shown to be divergent for any $\mu>0$ and, so, their practical use is doubtful.

\section{Particular cases and comparison with results from other deformed Bose-gas models}\label{sec3}

Consider two particular cases $r=2$ and $r=3$, earlier considered in~\cite{GR_EPJA} within a prescribed approximation. In these two cases, for the intercepts~$\lambda^{(2)}_\mu$ and~$\lambda^{(3)}_\mu$ ($0<\mu<\frac12$) we obtain:
\begin{align}
&\lambda^{(2)}_\mu = (1-e^{-\beta \hbar\omega})^{-1} \Bigl\{(1-e^{-\beta \hbar\omega})^{-1}
- \Bigl(\frac{1}{\mu}+\frac{1}{\mu^2}\Bigr) \Phi(e^{-\beta\hbar\omega },1,\mu^{-1})
- \Bigl(\frac{1}{\mu}-\frac{1}{\mu^2}\Bigr) \Phi(e^{-\beta\hbar\omega },1,\mu^{-1}\!-\!1)\Bigr\}\cdot \nonumber\\
&\qquad\Bigl((1-e^{-\beta \hbar\omega})^{-1} - \mu^{-1} \Phi(e^{-\beta\hbar\omega },1,\mu^{-1})\Bigr)^{-2} - 1,\label{lambda2}\\
&\lambda^{(3)}_\mu = (1\!-\!e^{-\beta \hbar\omega})^{-2} \Bigl\{(1\!-\!e^{-\beta \hbar\omega})^{-1}
\!-\! \Bigl(\frac{1}{\mu}\!+\!\frac{3}{2\mu^2}\!+\!\frac{1}{2\mu^3}\Bigr) \Phi(e^{-\beta\hbar\omega },1,\mu^{-1})
\!-\! \Bigl(\frac{1}{\mu}\!-\!\frac{1}{\mu^3}\Bigr) \Phi(e^{-\beta\hbar\omega },1,\mu^{-1}\!-\!1) - \nonumber\\
&\qquad- \Bigl(\frac{1}{\mu}\!-\!\frac{3}{2\mu^2}\!+\!\frac{1}{2\mu^3}\Bigr) \Phi(e^{-\beta\hbar\omega },1,\mu^{-1}\!-\!2)
\Bigr\}\cdot\Bigl((1\!-\!e^{-\beta \hbar\omega})^{-1} - \mu^{-1} \Phi(e^{-\beta\hbar\omega },1,\mu^{-1})\Bigr)^{-3} - 1.\label{lambda3}
\end{align}

It is worth also to consider the following specially constructed~\cite{Heinz_Zhang} function~$r^{(3)}_\mu(k)$ (also
considered in~\cite{Gavrilik_Sigma,GR_EPJA}), useful in the experimental context:
\begin{equation}\label{r^(3)}
r^{(3)}_\mu(k) = \frac{\lambda^{(3)}_\mu(k) - 3\lambda^{(2)}_\mu(k)}{2(\lambda^{(2)}_\mu(k))^{3/2}}.
\end{equation}
It implies cancellation of unwanted distortions and provides improved purity. With account of~(\ref{r^(3)}) and the exact expressions (\ref{lambda2})-(\ref{lambda3}) the explicit result for $r^{(3)}_\mu(k)$ readily follows merely by substitution (so we don't reproduce it here).

\begin{figure}[h]
\begin{minipage}[b]{0.45\linewidth}
\centering
\includegraphics[width = 1\linewidth]{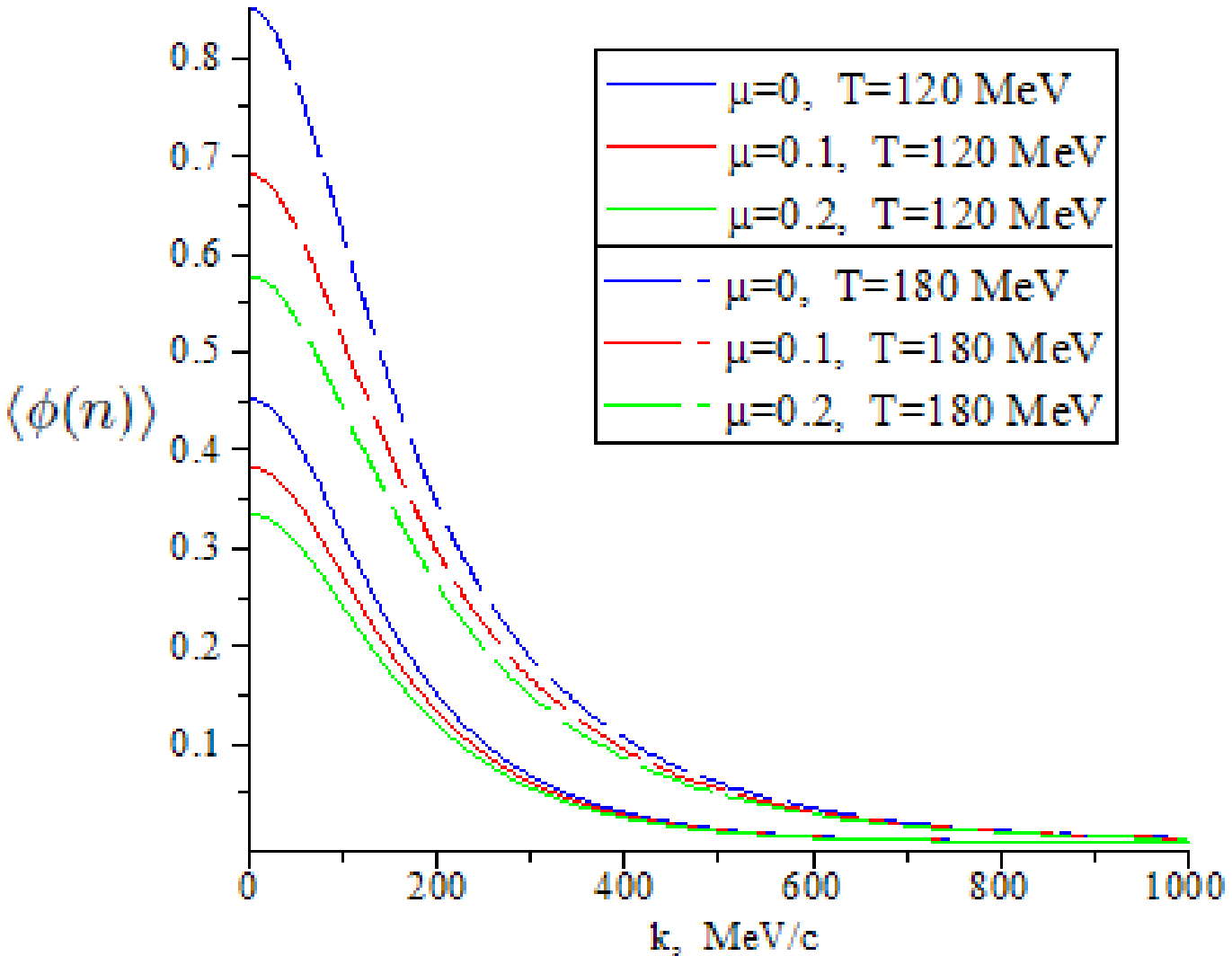}
\caption{\footnotesize Dependence of the deformed distribution $\langle\phi(N)\rangle = \langle a^\dag_k a_k\rangle$ on the momentum $k$, for deformation parameter values $\mu=0$ (pure Bose case) and $\mu=0.1,\,0.2$ (deformed case) and temperatures $T=120,\,180\,MeV$}
\label{fig1}
\end{minipage}
\hspace{0.5cm}
\begin{minipage}[b]{0.45\linewidth}
\centering
\includegraphics[width = 1\linewidth]{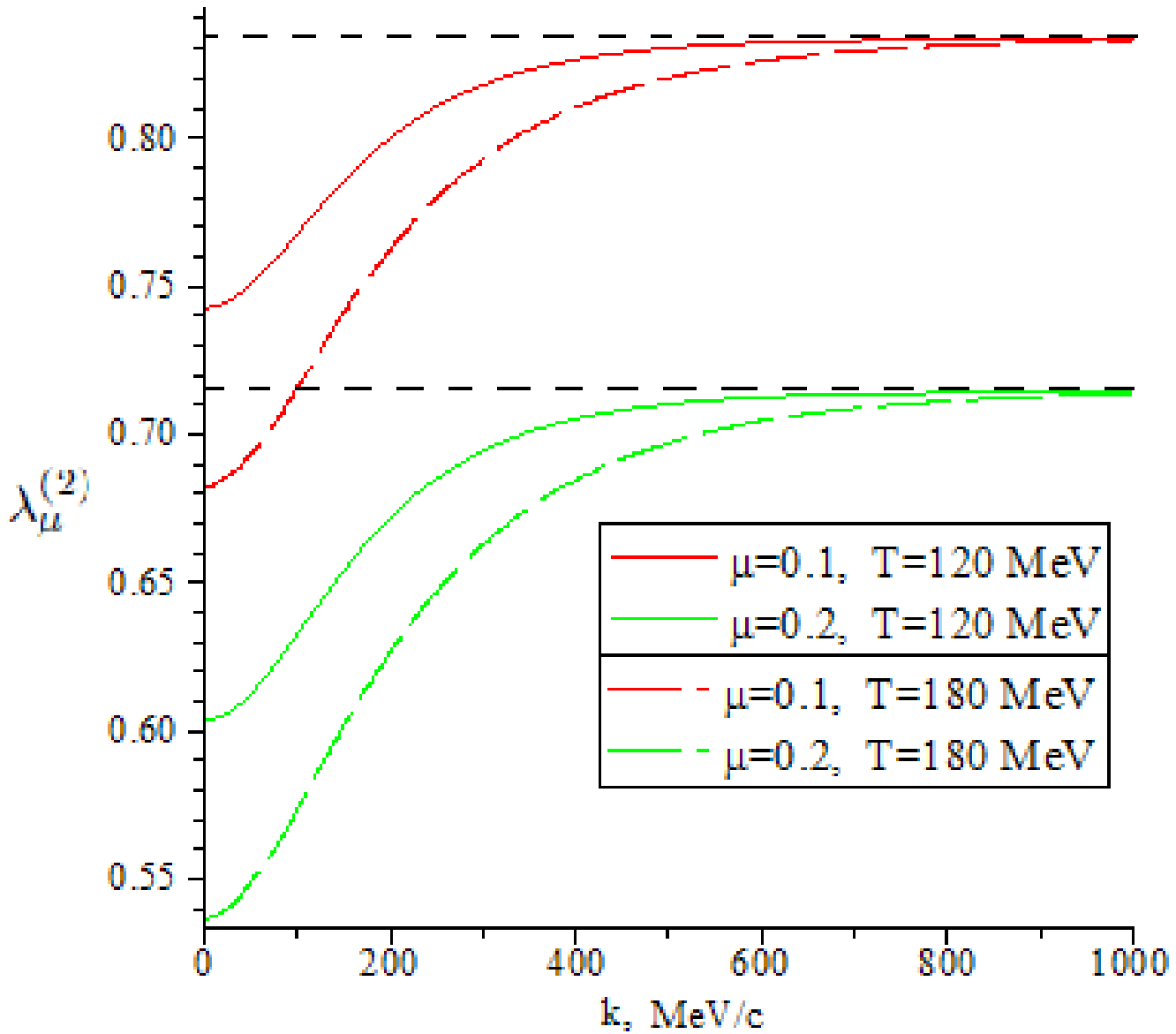}
\caption{\footnotesize Dependence of the intercept $\lambda^{(2)}_\mu(k)$ on the momentum $k$, for deformation parameter values $\mu=0.1,\,0.2$ and temperatures $T=120,\,180\,MeV$. Asymptotes are found from~(\ref{lambda23_asympt}) at $\mu=0.1,\,0.2$.
}
\label{fig2}
\end{minipage}
\end{figure}

\begin{figure}[h!]
\begin{minipage}[b]{0.45\linewidth}
\centering
\includegraphics[width = 1\linewidth]{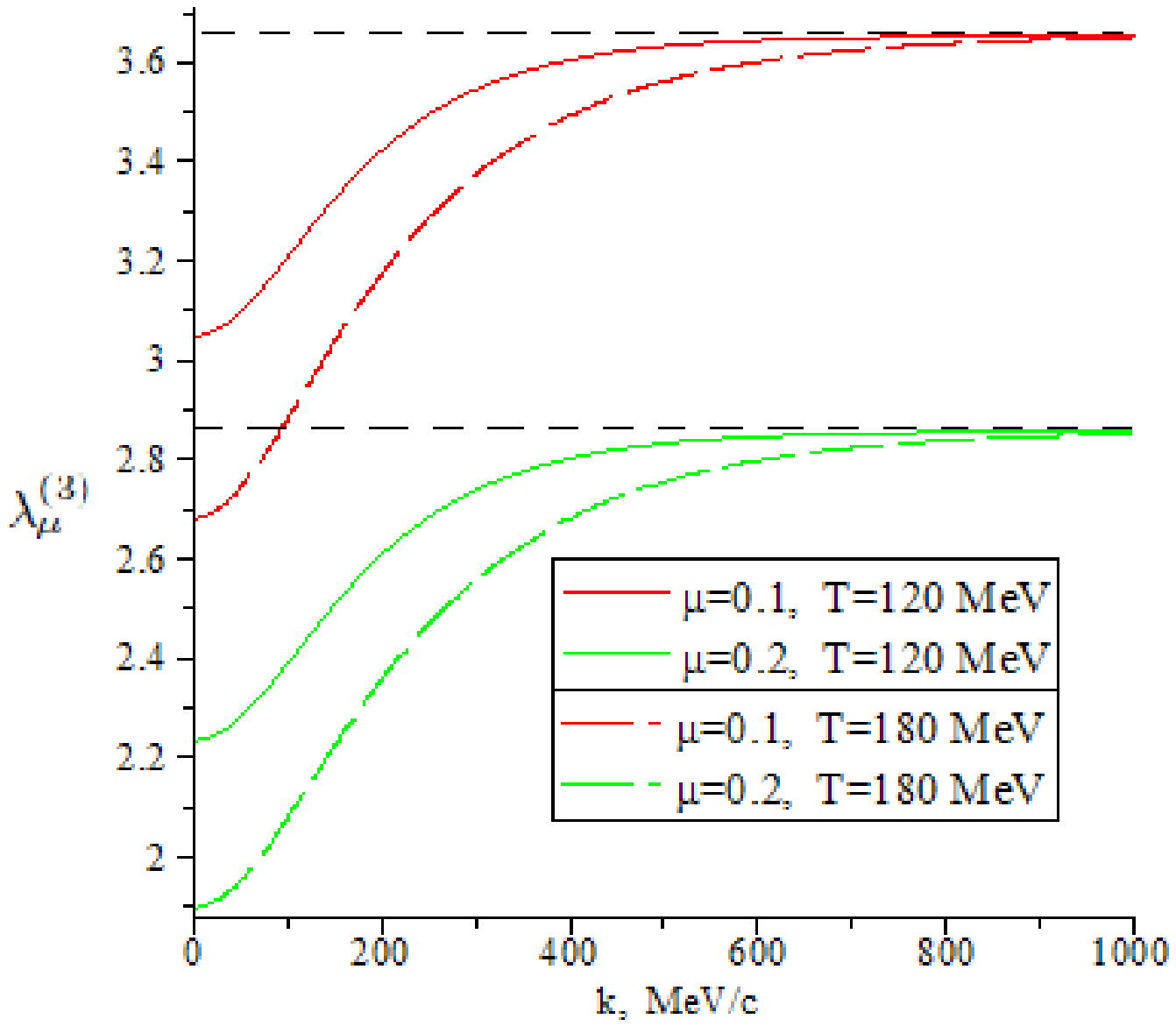}
\caption{\footnotesize Dependence of the intercept $\lambda^{(3)}_\mu(k)$ on the momentum $k$, for deformation parameter values $\mu=0.1,\,0.2$ and temperatures $T=120,\,180\,MeV$. Asymptotes are found from~(\ref{lambda23_asympt}) at $\mu=0.1,\,0.2$.
}
\label{fig3}
\end{minipage}
\hspace{0.5cm}
\begin{minipage}[b]{0.45\linewidth}
\centering
\includegraphics[width = 1\linewidth]{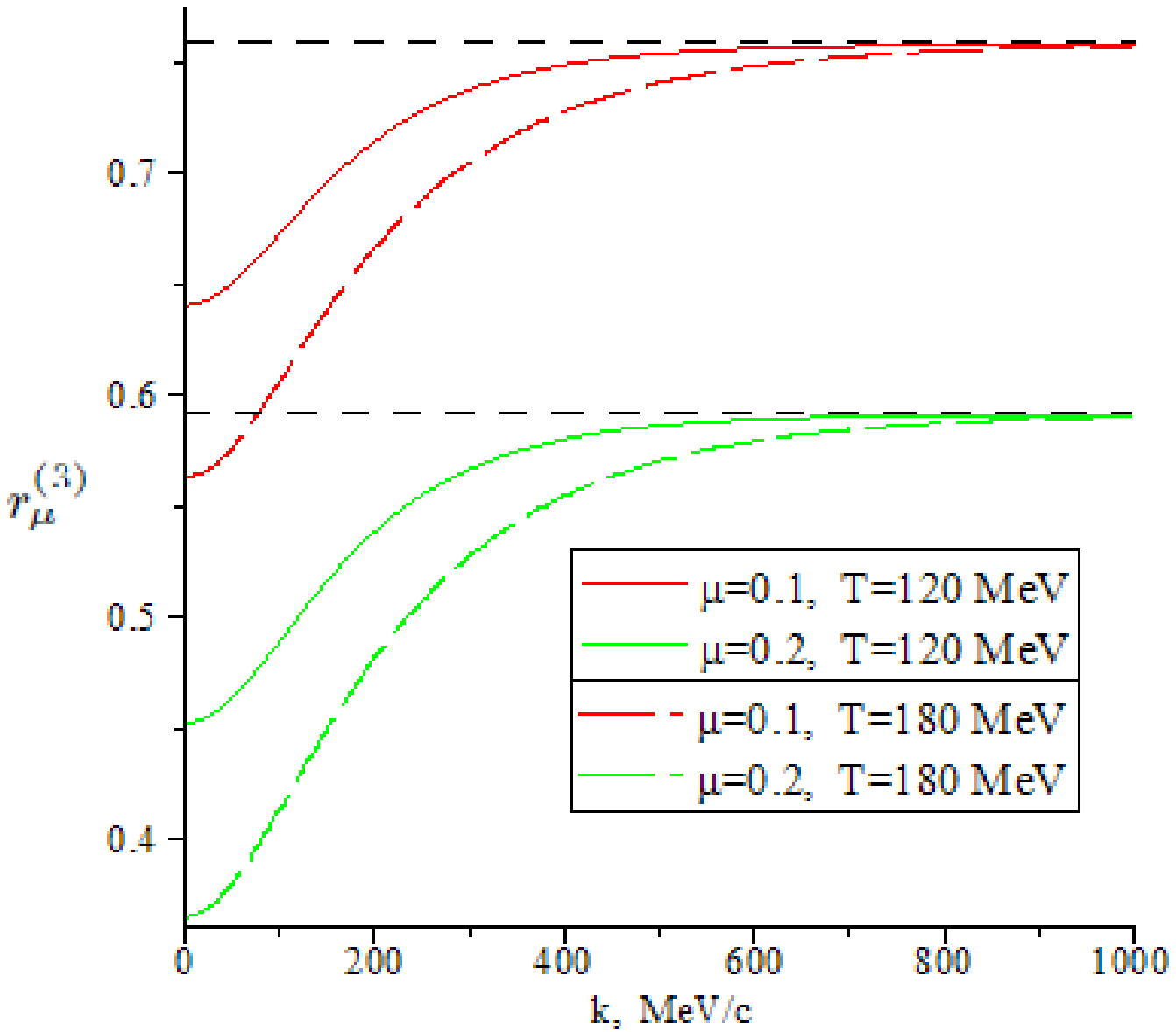}\\
\caption{\footnotesize Dependence of the correlation function $r^{(3)}_\mu(k)$ from~(\ref{r^(3)}) on the momentum $k$, for deformation parameter values $\mu=0.1,\,0.2$ and temperatures $T=120,\,180\,MeV$. Asymptotes stem from~(\ref{r^(3)}) after the substitution of~(\ref{lambda23_asympt}).}
\label{fig4}
\end{minipage}
\end{figure}

In case when deformed bosons (here of $\mu$-Bose gas) are used like in~\cite{Anchishkin2000,Gavrilik_Sigma,GR_EPJA} to describe relativistic particles, the energy is specified as $\hbar\omega_k=\sqrt{m^2+k^2}$. Then, the dependence of $\langle a^\dag_k a_k\rangle$, $\lambda^{(2)}_\mu(k,T)$, $\lambda^{(3)}_\mu(k,T)$ and $r^{(3)}_\mu(k,T)$ on the momentum~$k=|{\bf k}|$ for the values $\mu=0.1,\,0.2$ of deformation parameter, the temperatures being $T=120,\,180\,MeV$, is such as shown in Figures \ref{fig1}, \ref{fig2}, \ref{fig3} and \ref{fig4}.

From fig.~\ref{fig1} we see that the curves of deformed distributions lie below the usual ($\mu=0$) nondeformed Bose-Einstein one; with the increase of deformation parameter~$\mu$ or decrease of temperature $T$ the curves go lower and lower. For the momentum dependence of the intercepts~$\lambda^{(2)}_\mu(k)$, $\lambda^{(3)}_\mu(k)$, and the function~$r^{(3)}_\mu(k)$ (figures~\ref{fig2}-\ref{fig4}) we observe the similar behavior with respect to~$\mu$ -- larger deformation parameters~$\mu$ correspond to lower curves, while lower temperatures correspond to curves lying higher. Besides, for all the plots in figures~\ref{fig2}-\ref{fig4} the dependence on the momentum~$k$ shows asymptotical tending to the corresponding constant values for the intercepts (given by~(\ref{lambda23_asympt}) below), with the large momentum asymptotics for~$r^{(3)}_\mu(k)$ stemming from~(\ref{r^(3)}).

The remaining part of this section is devoted to a comparison of the results obtained above with the analogous results presented in earlier papers on other deformed Bose gas models. The comparison concerns the correlation intercepts along with their asymptotics found within $\mu$-Bose gas model~\cite{GR_EPJA}, and those found in deformed Bose gas model of another type. For instance, in the case of $p,q$-Bose gas model the exact expression for the deformed analog of $r$-particle distribution and for the intercept of the $r$th order correlation function obtained in~\cite{Adamska} respectively read (denote $[r]_{p,q}! \equiv [r]_{p,q}[r-1]_{p,q}...[1]_{p,q}$):
\begin{align}
&\langle (A^\dag_k)^r (A_k)^r\rangle = \frac{[r]_{pq}! (e^{\hbar\beta\omega}-1)}{\prod_{j=0}^r (e^{\hbar\beta\omega-p^j q^{r-j}})},\label{aa_dag_pq}\\
&\lambda^{(r)}_{p,q} (k) = \frac{\langle (A^\dag_k)^r (A_k)^r  \rangle}{\langle A^\dag_k A_k\rangle^r} -1 = [r]_{p,q}! \frac{(e^{\hbar\omega}-p)^r(e^{\hbar\omega}-q)^r}{(e^{\hbar\omega}-1)^{r-1} \prod\nolimits_{j=0}^r (e^{\hbar\omega}-q^{r-j}p^j)} - 1.\label{lambda_pq}
\end{align}
With respect to the corresponding result~(\ref{lambda_r}) of the present paper, the expression~(\ref{lambda_pq}), even though involving two parameters of deformation, has a simpler form (expressed in terms of elementary functions, not special ones), and this is presumably connected with the fact that $p,q$-oscillator belongs to the Fibonacci class whereas $\mu$-oscillator represents the class of quasi-Fibonacci~\cite{Gavrilik_QuasiFibonacci} oscillators (therefore its treatment is more involved).

Now let us examine the asymptotic behavior. It is seen that the correlation intercepts tend to certain constant values at large momenta. These asymptotical values of $\lambda^{(r)}_\mu(k,T)$ depend on $\mu$ only, and can be obtained directly from the definition~(\ref{lambda_def3}) by taking the limit
\begin{align}
\lambda^{(r)}_{\mu,\,asympt} = &\lim_{\omega\to\infty} \frac{\sum_{n=0}^\infty \frac{n}{1+\mu n}\cdot...\cdot\frac{n-r+1}{1+\mu (n-r+1)} e^{-\beta \hbar\omega n}}{(1-e^{-\beta\hbar\omega})^{r-1}\Bigl(\sum_{n=0}^\infty \frac{n}{1+\mu n} e^{-\beta \hbar\omega n}\Bigr)^r} - 1
= \lim_{\omega\to\infty} \frac{[r]_\mu ! e^{-\beta \hbar\omega r} + ...}{\bigl(\frac{1}{1+\mu}\bigr)^r e^{-\beta \hbar\omega r} + ...} - 1 =\nonumber\\
= &\bigl(1+\mu\bigr)^r [r]_\mu !-1, \qquad\qquad [r]_{\mu}! \equiv [r]_{\mu}[r-1]_{\mu}...[1]_{\mu}. \label{lambda_asympt}
\end{align}
In the numerator, the dominating $n=r$ term is retained whereas in the denominator the $n=1$ term dominates.
For $r=2$ and $r=3$ this result is in complete agreement with the corresponding asymptotical values of the $\mu$-Bose gas intercepts $\lambda^{(2)}$ and $\lambda^{(3)}$ earlier found in~\cite{GR_EPJA}:
\begin{equation}\label{lambda23_asympt}
\lambda^{(2)}_{\mu,\,asympt} = \bigl(1+\mu\bigr)^2 [2]_\mu!-1 = \frac{1}{1+2\mu}, \qquad \lambda^{(3)}_{\mu,\,asympt} = \bigl(1+\mu\bigr)^3 [3]_\mu!-1 = \frac{5+7\mu}{(1+2\mu)(1+3\mu)}.
\end{equation}
It is instructive to compare the asymptotical values~(\ref{lambda_asympt}) of the $r$th order correlation intercepts for $\mu$-Bose gas model with the corresponding asymptotics in the case of $p,q$-Bose gas model~\cite{Adamska,Gavrilik_Sigma}, that is
\begin{equation}\label{lambda_r_pq}
\lambda^{(r)}_{pq,\,asympt} = [r]_{p,q}!-1.
\end{equation}
The distinction, expressed on the $\mu$-Bose gas side by the extra factor~$\bigl(1+\mu\bigr)^r$, is connected with the fact that for $p,q$-oscillator we have $[1]_{p,q}=1$, whereas for $\mu$-oscillator $[1]_\mu=\frac{1}{1+\mu}\ne 1$. The exact expression for the asymptotics of the function~$r^{(3)}_\mu(k)$ can be obtained like in~\cite{GR_EPJA} by substitution of~(\ref{lambda23_asympt}) into (\ref{r^(3)}).

\section{Concluding remarks}\label{sec4}

This work describes further steps with respect to~\cite{GR_EPJA} -- here we extend those results and find them in precise form. More specifically, within $\mu$-Bose gas model we have obtained the {\it exact} expressions~(\ref{lambda_r}), (\ref{lambda_r_0}) for the intercepts of $r$th order correlation functions for any $r\ge 2$, as well as the explicit formulae (\ref{a^dag_a})-(\ref{mu=0_case}) for (deformed and non-deformed) one- and $r$-particle distributions. Recall that in~\cite{GR_EPJA} the corresponding formulae were given only for the cases $r=2$ and $r=3$, and within certain approximation. Here the exact expressions for these particular cases ($r=2$ and $r=3$) are presented in~(\ref{lambda2}) and (\ref{lambda3}) above. We have also obtained exact formula which gives the asymptotical values for the $r$th order correlation functions intercepts, as the corresponding functions of~$\mu$. Taylor expansions in the deformation parameter~$\mu$ for one- and $r$-particle (deformed) distributions are obtained, see~(\ref{num_expansion}) and (\ref{denum_expansion}). Further analysis shows that they turn out to be divergent for any $\mu>0$.

The dependence on the momentum $k$ of thermal average $\langle a^\dag_k a_k\rangle$, of the second and third order intercepts $\lambda^{(2)}_\mu(k)$ and $\lambda^{(3)}_\mu(k)$, as well as of the function $r^{(3)}_\mu(k)$, see~(\ref{r^(3)}), was illustrated in the corresponding figures~\ref{fig1}-\ref{fig4}. The comparison of the obtained results with those of previous papers shows that the plots of $\lambda^{(2)}_\mu(k)$, $\lambda^{(3)}_\mu(k)$ and $r^{(3)}_\mu(k)$ are in qualitative agreement with earlier results. Moreover, the corresponding large momentum asymptotics of
$\lambda^{(2)}_\mu(k)$ and $\lambda^{(3)}_\mu(k)$ completely agree with those found in~\cite{GR_EPJA}. If we compare the asymptotics~(\ref{lambda_asympt}) with the analogous asymptotics~(\ref{lambda_r_pq}) found, {\it also for arbitrary order $r$}, within the $p,q$-deformation~\cite{Adamska} we observe the following. In the both cases the corresponding deformation of $r$-factorial does appear, but the result~(\ref{lambda_asympt}) differs from the formula~(\ref{lambda_r_pq}), see also~\cite{Adamska}, by presence of the extra factor $(1+\mu)^r$ -- this is caused by differing values of the deformation structure functions at $n=1$.
At last, few words concerning the further research on the relevant issues. For more comprehensive treatment of the $\mu$-Bose gas model studied here and in~\cite{GR_EPJA} it is desirable to calculate some thermodynamic quantities (e.g. the critical temperature of condensation, the equation of state in the form of virial expansion) for $\mu$-Bose gas, like this was done in~\cite{Gavrilik_Virial} for the $p,q$-Bose gas model.

\section*{Acknowledgements}

The authors are thankful to A.P. Rebesh for useful discussions. This research was partially supported by the Special Program of the Division of Physics and Astronomy of the NAS of Ukraine.

\bibliographystyle{apsrev4-1}
\bibliography{references}

\end{document}